# The MATRIX: A Novel Controller for Musical Expression

**Dan Overholt**

MIT Media Laboratory
20 Ames St., Cambridge, MA 02139 USA
dano@media.mit.edu

## ABSTRACT
The MATRIX (Multipurpose Array of Tactile Rods for Interactive eXpression) is a new musical interface for amateurs and professionals alike. It gives users a 3-dimensional tangible interface to control music using their hands, and can be used in conjunction with a traditional musical instrument and a microphone, or as a stand-alone gestural input device. The surface of the MATRIX acts as a real-time interface that can manipulate the parameters of a synthesis engine or effect algorithm in response to a performer's expressive gestures. One example is to have the rods of the MATRIX control the individual grains of a granular synthesizer, thereby "sonically sculpting" the microstructure of a sound. In this way, the MATRIX provides an intuitive method of manipulating sound with a very high level of real-time control.

## INTRODUCTION
The use of computers today as part of the musical experience has allowed inventors and designers to create new types of controllers for musical expression. The goal of this work is to develop the MATRIX, an instrument that gives musicians and aspiring musicians a novel interactive performance and composition environment. While it is impossible to expect everyone to become proficient at composing music or playing an instrument, it is the author's hope that interfaces such as the MATRIX will engage people in new and active relationships with music.

The MATRIX is a tactile interface that controls a multi-modal musical instrument. It functions as a versatile controller that can be adapted to new tasks and to the preferences of individual performers in a variety of different environments. The interface provides musicians with a sculptable 3-D surface to control sound and music through a bed of push rods (see figure 1). Based on a user's real-time input, musical mapping algorithms control different types of signal processing algorithms.

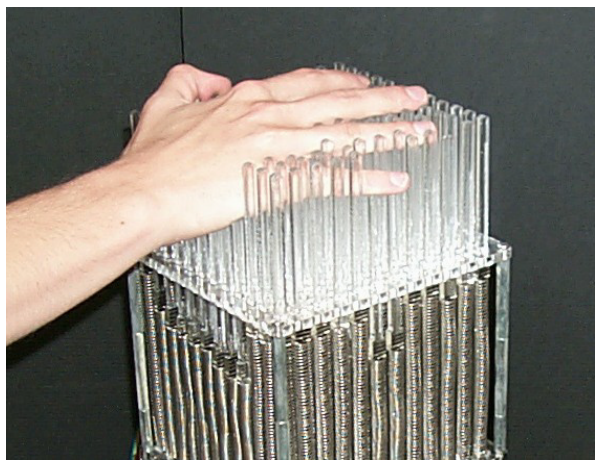

Figure 1 User interacting with the MATRIX interface.

In this paper I give a conceptual and technical overview of the MATRIX, including the hardware design and the software implementation for a variety of musical mappings. I begin with a discussion of the MATRIX at the system level, covering the design criteria, design approach, physical interface, and musical applications from a practical point of view. Finally I discuss user feedback and future directions for the project.

## RELATED WORK
The MATRIX results from work in the Hyperinstruments and Interactive Cinema groups at the MIT Media Laboratory [9]. Many interactive performance-driven instruments came out of the work of Tod Machover, who coined the term 'Hyperinstrument' to describe a number of the interfaces that were developed for projects such as the Brain Opera [5]. Also from the Hyperinstruments group, Gil Weinberg and Seum-Lim Gan have created a series of musical instruments called squeezables [4,14], which capture the gestures of squeezing and pulling foam, gel and plastic balls, and Teresa Marrin built a conductor's jacket that maps gestures to musical intentions [6]. Finally, in the Responsive Environments group, Dr. Joe Paradiso has created a number of musical interfaces, some of which include the Expressive Footwear project [11], and Musical Trinkets [10] that use a swept-frequency tag reading interface.

Many others have created gestural musical interfaces. Some examples related to this work include controllers developed by Don Buchla [2] such as Thunder, which is a specialized MIDI controller that senses various aspects of the touch of hands on its playing surface, and transmits the resultant gestural information to responsive electronic instrumentation. In industry, Tactex Controls, Inc. is currently producing one of the only commercially available multiple touch capable control surfaces [13]. It utilizes fiber optic-based Kinotex® technology, which was developed by Canpolar East, Inc. for the Canadian Space Agency. Another company that is developing multiple touch interfaces is FingerWorks [3], which is in the process of developing a product called the FingerBoard that will use their GestureScan technology to replace a standard keyboard and mouse with a multi-touch interface. While all of these interfaces allow multiple points of interaction, they are sensitive only to two-dimensional pressure, not three-dimensional motion like the MATRIX.

**INTERFACE CONCEPT AND DESIGN**

The human hand has shaped and produced many amazing things over the centuries—it is one of the most nimble and dexterous tools at our disposal. Thousands of musical instruments have been developed that are played with the hand, and one of the goals of the MATRIX is to create an interface that will make the most of our skills with our hands. The MATRIX is an example of a human-computer interface that can extract useful information from a human gesture, exploiting the senses of touch and kinesthesia. It can respond in real-time to the shape of a three-dimensional surface, and attempts to interpret this in a musically meaningful way. The interface captures the movement of the hand with as much resolution as possible, thus taking advantage of the skills we have developed through a lifetime of interaction with the physical world.

There are several ways in which the MATRIX is optimized in this respect—the most obvious being the physical size and shape of the interface, which is based around the structure of the hand. The design uses an array of rods in which each is roughly the size of a human finger, and the entire array is slightly larger than an average hand. Having multiple points of interaction in this way allows for a variety of playing techniques, ranging from the smooth, wave-like continuous control of the shape of the surface (using the open palm and fingers), to the individual control of single rods (or a group of rods) using the fingertips. By using a bed of rods to provide parallel input specification for such a large number of control points, the MATRIX improves the communication bandwidth with the computer, thereby giving musicians a much greater potential for control.

**HARDWARE IMPLEMENTATION**

The MATRIX has a total of 144 rods in a 12 by 12 grid, with a density of about 4 rods per square inch, covering an area of 36 square inches. The prototype is made of clear plexiglass (acrylic), and consists of a top plate, a bottom plate, 144 spring-mounted rods, and thirteen circuit boards. All of the plexiglass parts are made using a laser-cutter. The rods move vertically within their guides, and are held up by springs at rest. The range of travel for each rod is approximately 4 inches.

The position of each rod is determined using opto-electronics, and there are twelve MATRIX optics boards—one for each row of 12 rods. The sensing technique used is quadrature encoding, and is similar to the technique used in shaft-encoders. It uses offset infrared transmitter/receiver pairs and a moiré pattern on the rods to allow the sensing architecture to derive the direction and rate of travel for each rod (see figure 2). In order to decode this quadrature signal, a state machine determines which direction the rod has moved, and a bank of 7-bit counters keeps track of each rod's current position. It is assumed that the rods are at position zero when the system is booted, and the counters are bounded at each end to keep them from wrapping around. In this way, it is possible to re-calibrate the system while in operation by simply pushing the rods through their entire range of motion.

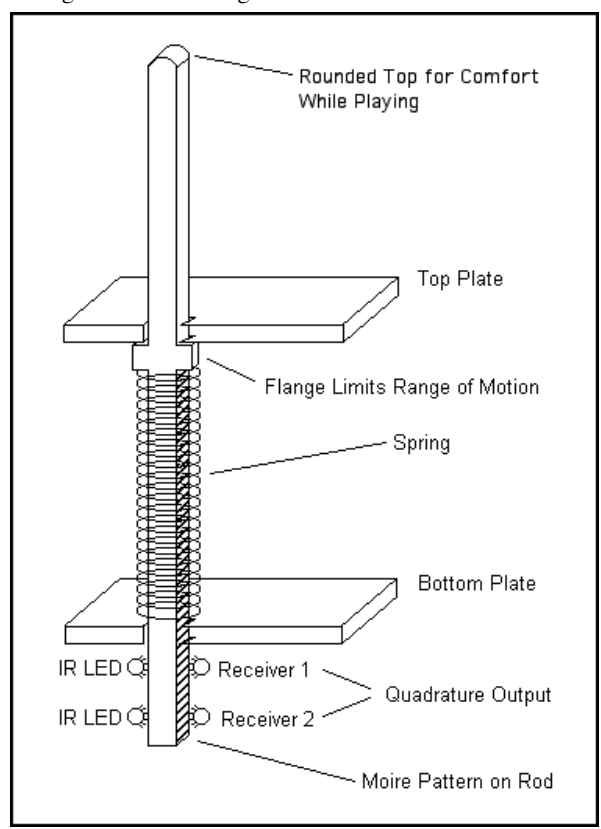

Figure 2 Diagram of a MATRIX rod.

The main MATRIX circuit board takes the signals from the twelve optics boards and calculates the 3-D topology of the surface using an FPGA (Field Programmable Gate Array). All of the resulting positions are continually sent to a host computer at 57.6 kbaud serial rate in order to achieve an overall frame rate of about 30 Hz. This lets the system respond to a user's input with no perceptible delay, allowing for smooth, continuous changes in the resulting audio. In addition to the FPGA (which is programmed in VHDL [1]), the main MATRIX circuit board has an EPROM to boot the FPGA, a PIC microcontroller, and a serial line driver to transmit the data. The overall size of the main circuit board is designed to match the size of the top and bottom plates through which the rods are guided. This allows the vertical supports for the MATRIX to hold the circuit board in place along with all of the acrylic parts (see figure 3).

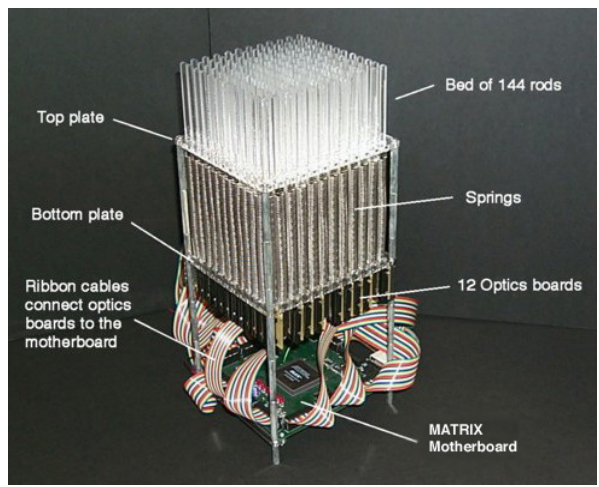

Figure 3 MATRIX hardware components.

## MUSICAL APPLICATIONS

Musical instruments have traditionally been divided according to their technique of producing sound (strings, winds, percussion, etc.), but the MATRIX does not fit into any of these categories. Instead, it allows for many different types of musical interactions, which makes it an adaptive musical instrument. By making the interface flexible, individuals can configure the instrument to respond best to their skill level and style of playing. The MATRIX is an attempt at creating a dynamic musical instrument that will acknowledge the natural methods of human expression, and allow people to improve their skill and knowledge in the course of musical discovery. The MATRIX provides the potential to explore a new approach to adaptable learning with a unified interface in which each level leads to the next. Inexperienced users could first experiment with algorithmic music generation, allowing them to engage in musical expression without control of specific notes, harmonies, or timbres (the mapping would not immediately demand virtuosi levels of skill and precision). Then using the same interface, users could proceed to learn about timbres, dynamics, and sound modification, exploring more complex composition, performance, and sound shaping techniques.

In this section I describe three categories of musical applications that provide different levels of musical control suitable for the adaptable learning approach described above: a direct synthesis mode (low-level audio shaping), a signal processing mode that allows the MATRIX to shape the sound of a traditional musical instrument, and a higher-level gestural mode that allows users to algorithmically generate music using only the MATRIX. Some preliminary musical applications have been explored for each of these categories[1]. Finally I discuss the remote and distributed operation modes that allow the MATRIX to act as a controller for collaborative performance.

### Direct Synthesis

The direct synthesis mode is designed to allow people to use the MATRIX for the creation of original sounds by directly controlling the output of a software synthesizer. One method of achieving this is to use the shape of the MATRIX's surface to control the harmonic spectrum of an additive synthesizer—something the interface is particularly well suited for, as it provides an intuitive mapping of one harmonic for each rod. A performer or composer would have to develop considerable skill in controlling the MATRIX in order to gain mastery of the sounds produced using such direct mappings.

A preliminary additive synthesis algorithm was implemented in Max/MSP [7] on an early prototype of the MATRIX that had only 12 functioning rods. The algorithm uses a bank of sinusoid oscillators to generate audio. The synthesis can also be performed in SuperCollider [12] using an inverse FFT (Fast Fourier Transform), which is more efficient for larger numbers of harmonics.

### Signal Processing

In the signal processing mode, the MATRIX allows users accustomed to a traditional instrument to modify its sound through various effects algorithms. By allowing the user to process a signal of their own creation, the MATRIX becomes more than a synthesizer. It gives a performer the ability to modify a live sound (i.e., through a microphone), or play an instrument such as a keyboard with one hand while shaping the sound with their other hand (as a "left-hand controller"). This type of interaction with the MATRIX requires at least some musical skill and understanding in order to produce interesting results.

A real-time granular synthesis algorithm has been implemented in SuperCollider. The algorithm takes live audio, breaks it into individual 'grains' of sound, and assigns each sonic grain to a particular rod on the MATRIX. It is intended to be used with a microphone, into which users sing or play an instrument and simultaneously modify the microstructure of a sound in real-time. The algorithm allows the user to change grain-length, modify level, pitch-shift, and re-order the grains of sound using the sculptable surface of the MATRIX.

### Algorithmic and Gestural Music

This mode allows novices to engage in musical expression without control of specific notes, harmonies, or timbres. Instead of requiring excessive musical knowledge and physical precision, performers will be able to think about music in terms of emotional outputs and the gestures that feel natural to express them. This should let a musician go much more directly from a musical idea or feeling to sound, thinking more about qualitative musical issues than technique or physical manipulations. Inexperienced musicians will likely enjoy this higher level of interaction, as it allows them create music by expressively manipulating the 3-D surface formed by the MATRIX's rods.

---

[1] Video and audio clips are posted at the following web site: http://www.media.mit.edu/~dano/matrix/.

The first example of such an application was created in Max on the same early prototype of the MATRIX as the additive synthesis algorithm. It mapped twelve of the MATRIX's rods to various drum sounds, depending on the activity level and excursion of the performer's gestures. Subtle, gentle movements resulted in soft, high-pitched drum sounds, while more vigorous movements caused louder, more diverse, and explosive sounds. This approach demonstrated a very basic mapping of gesture to musical intent. The application used MIDI to control an E-mu Audity 2000 sound module, which produced the drum sounds that were triggered by a user's interactions with the MATRIX.

### Remote and Distributed Operation

Through the use of OSC (Open Sound Control) [8], the MATRIX can be used as a device for remote interaction with a media performance. Preliminary tests have been done between MIT in Cambridge, Massachusetts and the Media Lab Europe in Dublin, Ireland. These tests were based on the additive synthesis algorithm, using Max to send OSC control data to SuperCollider via the Internet. Such a setup could for instance be used to modify the timbre of a remote performer's instrument during a live performance. With high-speed access on both ends, the response time was surprisingly good, with delays of much less than 0.5 seconds in almost all cases.

### USER FEEDBACK

The MATRIX has been demonstrated to the public twice (at two different Media Lab open houses), as well as to many of the Media Lab sponsors. User responses to the MATRIX were in general positive. The bed of rods on the MATRIX naturally affords pushing, much like the popular 'pin-screen' toy, so users generally needed no explanation for how to interact with the device. After having played with the MATRIX for a few moments, many users commented on how natural the interaction was. Most people seemed to especially enjoy modes in which the mappings from their gestures to the resulting music were easy to comprehend, while less direct mappings made some users wonder what they were affecting in the music. These observations have shown that it is necessary to ensure that users can easily correlate their actions to effects in the music.

### CONCLUSIONS AND FUTURE WORK

The simple musical demonstrations included here only hint at the new possibilities enabled by such a musical interface. Future work will explore more complicated mappings, as well as physical interface improvements. The force required to push the rods all the way down is a drawback with the current mechanical design. Newer versions of the MATRIX will include lighter springs that will allow the interaction to more easily use the full range of motion. Additionally, I will endeavor to find signal processing techniques that allow the most extensive control of an audio signal using the MATRIX's expressive abilities. Some of the methods to be investigated include wave terrain synthesis and sound spatialization algorithms. Finally, in fulfillment of the artistic goals of the MATRIX project, the interface will be used to compose and perform new music in concert settings.

### ACKNOWLEDGMENTS
I would like to thank Professors Tod Machover, Glorianna Davenport, and Joe Paradiso, as well as Ali Mazalek, Tristan Jehan, Paul Nemirovsky, and Andrew Yip for their help with this project.

### REFERENCES

1. Ashenden, P. The Designer's Guide to VHDL. Morgan Kaufmann, San Francisco, CA. 1995.

2. Buchla, D. home page, http://www.buchla.com/

3. FingerWorks, http://www.fingerworks.com/

4. Gan, S. Squeezables: Tactile and Expressive Interfaces for Children of All Ages. Cambridge, MIT Media Lab Masters Thesis, 1998.

5. Machover, T. Brain Opera Update, January 1996. Internal Document, MIT Media Laboratory, 1996.

6. Marrin, T. & Picard, R. The Conductor's Jacket: a Testbed for Research on Gestural and Affective Expression. Cambridge: MIT Media Lab.

7. Max/MSP – Cycling'74, Publisher of Max and MSP. http://www.cycling74.com/

8. Open Sound Control at U.C. Berkeley's CNMAT (Center for New Music and Audio Technologies), http://cnmat.cnmat.berkeley.edu/OSC/

9. Overholt, D. The Emonator: A Novel Musical Instrument. MIT Media Laboratory Master's Thesis, 2000.

10. Paradiso, J. & Hsiao, K. Musical Trinkets: New Pieces to Play, in SIGGRAPH 2000 Conference Abstracts and Applications, ACM Press, NY, July 2000, p. 90.

11. Paradiso, J., Hsiao, K., & Hu, E. "Interactive Music for Instrumented Dancing Shoes." Proc. of the 1999 International Computer Music Conference, October 1999, pp. 453-456.

12. SuperCollider, http://www.audiosynth.com/

13. Tactex Controls, Inc. Multi-touch, pressure sensitive input devices. Tactex web site: http://www.tactex.com/

14. Weinberg, G. Expressive Digital Musical Instruments for Children. Cambridge, MIT Media Lab Master's Thesis, 1999.